\documentclass{PoS}

\usepackage{etex, xy,cite}
\xyoption{all}

\usepackage{amssymb, amsbsy, amsfonts}
\usepackage{amsmath, amsthm}

\newcommand{\NN}{\mathbb{N}}
\newcommand{\ZZ}{\mathbb{Z}}
\newcommand{\QQ}{\mathbb{Q}}
\newcommand{\KK}{\mathbb{K}}

\newcommand{\ep}{\varepsilon}

\newtheorem{remark}{Remark}

\newcommand{\vect}[1]{{\bf #1}}

\newcounter{linectr}
\newenvironment{myEnumerate}{\begin{list}{(\arabic{linectr})}{\usecounter{linectr}
\labelwidth1ex\itemsep0ex\labelsep1ex\leftmargin2ex\parskip0.0cm\topskip0cm\partopsep0cm
\listparindent0ex}}{\end{list}}

\newcounter{linectr2}
\newenvironment{myEnumerate2}{\begin{list}{(\alph{linectr2})}{\usecounter{linectr2}
\labelwidth1ex\itemsep0ex\labelsep1ex\leftmargin2ex\parskip0.0cm\topskip0cm\partopsep0cm
\listparindent0ex}}{\end{list}}

\title{{\footnotesize DESY 16-160, DO-TH 16--19}\\ 
Algorithms to solve coupled systems of differential equations in terms of power series\thanks{This 
work
was supported in part by the Austrian Science Fund (FWF) grant SFB F50 
(F5009-N15) and the European
Commission through contract PITN-GA-2012-316704 ({HIGGSTOOLS}).}}

\ShortTitle{Algorithms to solve coupled differential systems in terms of power series}

\author{Jakob Ablinger and \speaker{Carsten Schneider}
\\
        Research Institute for Symbolic Computation (RISC)\\
        Johannes Kepler University, Altenbergerstra\ss{}e 69, A-4040 Linz,
Austria\\
        E-mail: \email{Jakob Ablinger@risc.jku.at, Carsten.Schneider@risc.jku.at}}

\author{Arnd Behring, Johannes Bl\"umlein and Abilio de Freitas\\
        Deutsches Elektronen--Synchrotron, DESY,\\
Platanenallee 6, D--15738 Zeuthen, Germany\\
        E-mail: \email{johannes.bluemlein@desy.de, abilio.de.freitas@desy.de, arnd.behring@desy.de}}
\abstract{Using integration by parts relations, Feynman integrals can be represented in terms of 
coupled systems of differential equations. In the following we suppose that the unknown Feynman integrals 
can be given in power series representations, and that sufficiently many initial values of the integrals are given. Then there exist algorithms that decide constructively if the coefficients of their power series representations can be given within the class of nested sums over hypergeometric products. In this article we will work out the calculation steps that solve this problem. First, we will present a successful tactic that has been applied recently to challenging problems coming from massive 3-loop Feynman integrals. Here our main tool is to solve scalar linear recurrences within the class of nested sums over hypergeometric products. Second, we will present a new variation of this tactic which relies on more involved summation technologies but succeeds in reducing the problem to solve scalar recurrences with lower recurrence orders. 
The article will work out the different challenges of this new tactic and demonstrates how they can be treated efficiently with our existing summation technologies.}

\FullConference{Loops and Legs in Quantum Field Theory\\
		24-29 April 2016\\
		Leipzig, Germany}

\begin{document}

%---------------------------------------------------------------------------------------------------
\section{Introduction}
%---------------------------------------------------------------------------------------------------

\vspace{1mm}
\noindent
Coupled systems of linear differential equations arise frequently in intermediate calculations 
when one tries to tackle massless or massive three-loop Feynman integrals. Namely, using 
integration 
by parts (IBP) methods\footnote{In our calculations we used heavily the {\tt C++} program 
{\tt Reduze~2}~\cite{Reduze2} based on Laporta's algorithm \cite{Laporta:2001dd}.}~\cite{IBP} one 
can determine such a coupled system and together with the initial values one obtains a complete 
description of the involved Feynman integrals. Various techniques~\cite{DEQ} have been elaborated 
to extract relevant information from these systems that sheds light on the involved Feynman 
integrals. For heuristic methods to find closed form solutions of such systems, in case the 
solutions are given by iterative integrals, see for 
instance~\cite{Kotikov:2010gf}.
A completely algorithmic approach has been worked out 
recently in~\cite{NewUncouplingMethod,VLadders,UncoupleTHM}
if the unknown functions $\hat{I}_i(x)$ with $1\leq i\leq r$ of the coupled system admit 
power series representations\footnote{In addition there may be trailing terms such as $\ln^k(x)$ 
for each of these power series and one may shift $N$ by a finite number of integer values.}
\begin{equation}\label{Equ:HatIAnsatz}
\hat{I}_i(x)=\sum_{N=0}^{\infty}I_i(N)x^N.
\end{equation}
Then under the assumption that sufficiently many initial values $I_i(\nu)$ at nonnegative integers $\nu\in\NN$ are available, we can utilize our summation machinery that decides constructively if the $I_i(N)$ can be written in terms of nested sums over hypergeometric products. This means that they can be composed by elements from the rational function field $\KK(N)$, the three operations 
($+,-,\cdot$), hypergeometric products of the form $\prod_{k=l}^Nh(k)$ with $l\in\NN$ and $h(k)$ being a rational function in $k$ and being free of $N$, and sums of the form $\sum_{k=l}^Nh(k)$ with $l\in\NN$ and with 
$h(k)$ being a nested hypergeometric sum expression w.r.t.\ $k$ and being free of $N$. This class of special functions covers as special cases harmonic sums~\cite{BlumVerm}, generalized harmonic sums~\cite{Moch:2001zr,Ablinger:2013cf}, cyclotomic 
harmonic sums~\cite{Ablinger:2011te} or nested binomial 
sums~\cite{BinSums,Ablinger:2014bra}; for surveys on these quantities see e.g. \cite{REF1}. 

The central step of our procedure~\cite{NewUncouplingMethod,VLadders} is to translate the problem 
of solving a coupled system of differential equations to the problem of solving scalar recurrences 
that 
depend only on one of the unknown functions $I_i(N)$. To obtain a complete algorithm we utilize
uncoupling algorithms~\cite{UNCOUPL,Zuercher:94}, recurrence solving algorithms~\cite{dAlembert,ParticularSol,BKSS:12}, and simplifiers of the solutions by means of symbolic summation~\cite{Summation} based on difference field~\cite{DFTheory} and difference ring~\cite{DRTheory} theory. 
This successful interplay of the underlying summation tools (uncoupling, recurrence solving, and 
indefinite summation) has been implemented within the package \texttt{SolveCoupledSystem}~\cite{UncoupleTHM} that relies on the summation packages~\texttt{Sigma} and~\texttt{SumProduction}~\cite{Summation} (concerning the summation tools) and the the package~\texttt{OreSys}~\cite{OreSys} (executing Z\"urcher's uncoupling algorithm~\cite{Zuercher:94}). The package \texttt{SolveCoupledSystem} has been applied successfully for various challenging 3-loop calculations~\cite{DiffCalculations}. These heavy calculations were possible by incorporating besides \texttt{Sigma} also the package~\texttt{HarmonicSums}: whenever possible, we employed its algorithms that are tuned for harmonic sums, cyclotomic sums, generalized harmonic sums and nested binomial sums.

Still, in recent calculations two serious problems arose. First, in some instances we obtained 
coupled systems of difference equations that we failed to uncouple due to restricted time and 
memory resources. Second, we succeeded in uncoupling the system but derived scalar recurrences 
whose orders were very large. We emphasize that \texttt{Sigma} was still capable of solving these 
large recurrences (in~\cite{GuessingProject} we solved gigantic recurrences up to order 35) but 
finding 16 or more initial values to combine the solutions accordingly was the bottleneck: 
extracting such a large amount of initial values from the given Feynman integrals was out of 
the scope 
of the existing algorithms, like exploiting the $\alpha$-parameterization of the 
integrals~\cite{Ablinger:2014uka}.\footnote{Yet one might think of other methods to obtain these 
initial values. It is, however, clear that the calculation of finite Mellin moments of complicated 
diagrams in general constitutes no simple task either, cf.~\cite{MOM}, for very high moments.}
Based on our existing algorithmic machinery (see Subsection~\ref{Sec:PureREC} below) we will present a new tactic in Subsection~\ref{Sec:DEREC} that has been implemented within the package \texttt{SolveCoupledSystem} and that overcomes the problems described above. A conclusion concerning future calculations that will rely on this new machinery will be given in Section~\ref{Sec:Conclusion}.
The present algorithm also decides, whether or not a system can be solved in terms of iterative 
integrals over whatsoever alphabet or not, and allows therefore to single out systems which are 
not 
first order reducible requiring other techniques of solution, cf.~\cite{CIS}.

%-----------------------------------------------------------------------------------------------------
\section{Two basic strategies}
%-----------------------------------------------------------------------------------------------------

\vspace{1mm}
\noindent
Suppose we are given a coupled system of differential equations coming from IBP-methods~\cite{IBP,Laporta:2001dd,Reduze2}. More precisely, we are given a matrix $\hat{A}$ of dimension $r\times r$ with entries from the rational function field $\KK(x)$ and we are given a vector $\vect{\hat{b}}(x)=(\hat{b}_1(x),\dots,\hat{b}_r(x))$ of length $r$ where each entry is given in terms of a linear combination of master integrals with coefficients from $\KK(x)$.
Then, given sufficiently many initial values we seek for the uniquely determined solution 
$\vect{\hat{I}}(x)=(\hat{I}_1(x),\dots,\hat{I}_r(x))$ 
of the system
\begin{equation}\label{Equ:DEProblem}
D_x\vect{\hat{I}}(x)=\hat{A}\,\vect{\hat{I}}(x)+\vect{\hat{b}}(x)
\end{equation}
with $D_x\vect{\hat{I}}(x)=(D_x\hat{I}_1(x),\dots,D_x\hat{I}_r(x))$, where $D_x$ is the 
differential operator w.r.t.\ $x$ acting on functions in $x$. 

In our setting we may assume that the master integrals arising in the $\hat{b}_i(x)$ have a power series 
representation, where the coefficients are already computed. Within our calculations in~\cite{DiffCalculations} this has been accomplished with 
\begin{enumerate}
 \item the package \texttt{EvaluateMultiSums}~\cite{Summation} relying on the summation package \texttt{Sigma} and the special functions package \texttt{HarmonicSums}~\cite{HarmonicSums} (see also~\cite{BlumVerm,Moch:2001zr,Ablinger:2013cf,Ablinger:2011te,BinSums,Ablinger:2014bra});
 
 \item the package \texttt{MultiIntegrate}~\cite{Integration} being a specially tuned version of the multivariate Almkvist--Zeilberger algorithm;
 \item the package \texttt{SolveCoupledSystem} containing the algorithms that will be explained below -- here the package has been  applied to systems which do not depend on the functions given in $\vect{I}(x)$.
\end{enumerate}
In particular, we assume that the coefficients are given in terms of nested sums over hypergeometric products.
This enables one to write also the $\hat{b}_i(x)$ in a power series representation
\begin{equation}\label{Equ:HatbAnsatz}
\hat{b}_i(x)=\sum_{N=0}^{\infty}b_i(N)\,x^N
\end{equation}
whose coefficients $b_i(N)$ are given in terms of nested sums over hypergeometric products.
Furthermore, we assume that also the unknown master integrals $\hat{I}_i$ are analytic, i.e., there exist the power series representations~\eqref{Equ:HatIAnsatz}.\\
Under these assumptions we seek for the coefficients $I_1(N),\dots, I_r(N)$ in closed form. This means that we look for representations in terms of nested sums over hypergeometric products whenever such a representation is possible.

In Subsection~\ref{Sec:PureREC} we will recall our first 
algorithm~\cite{NewUncouplingMethod,VLadders,UncoupleTHM} that solves this problem and that has been successfully applied to various challenging massive 3-loop Feynman integrals~\cite{DiffCalculations}. In this approach we will address some efficiency problems that have occurred recently. By a variation of the employed building blocks, we will present a new tactic that overcomes these problems in Subsection~\ref{Sec:DEREC}.

%-------------------------------------------------------------------------------------------
\subsection{Tactic 1: Uncoupling a system of difference equations}\label{Sec:PureREC}
%-------------------------------------------------------------------------------------------

\vspace{1mm}
\noindent
In the first approach the system of differential equations~\eqref{Equ:DEProblem} with the unknown functions $\hat{I}_i(x)$, which have power series representations~\eqref{Equ:HatIAnsatz}, is translated to a system of difference equations for the corresponding coefficients $I_i(N)$. Afterwards one applies computer algebra algorithms in order to find a symbolic representation of the $I_i(N)$ in terms of nested sums over hypergeometric products. 
\noindent More precisely, we execute the following steps.

\begin{myEnumerate}
\item We utilize that an analytic function $\sum_{N=0}^{\infty}h(N)x^N$
with a certain convergence disc fulfills  
\begin{equation}\label{Equ:DRelation}
D_x\sum_{N=0}^{\infty}h(N)x^N=\sum_{N=1}^{\infty}N\,h(N)x^{N-1};
\end{equation}
similarly, if we argue in the formal power series setting, the operator $D_x$ is just defined in this way.
Plugging the already computed series expansions of $\hat{b}_i(x)$ and the Ansatz~\eqref{Equ:HatIAnsatz} into~\eqref{Equ:DEProblem} (note that in general we can start with a coupled system of differential equations of higher-order; see~\cite{UncoupleTHM}) 
and doing coefficient comparison w.r.t.\ $x^N$ yields a coupled system of linear difference 
equations, say of order $m\geq0$, which is of the form
\begin{equation}\label{Equ:CoupledDifferenceSys}
A_m\,\vect{I}(N+m)+A_{m-1}\,\vect{I}(N+m-1)+\dots+A_0\,\vect{I}(N)=\vect{b}(N),
\end{equation}
with $\vect{I}(N+l)=(I_1(N+l),\dots,I_r(N+l))$ for $l\in\NN$ and where the matrices $A_m,\dots,A_0$ have dimension $r\times r$ with entries from $\KK(n)$. Ideally, one should solve this coupled system directly. For rational solutions this problem is solved in~\cite{SolveCoupled}; for first steps towards the more general setting of $\Pi\Sigma$-fields we refer to~\cite{SolveCoupledPiSigma}. So far, there are no algorithms available to solve such systems directly in terms of nested sums over hypergeometric products. Therefore we will proceed as follows. 

\item We uncouple the system such that some of the unknown coefficients $I_i(N)$ of the power series are solutions of a scalar linear recurrence (which does not depend on other unknown functions) and 
where the remaining unknown coefficients can be given as a linear combination of the coefficients 
that are determined by the solutions of the scalar recurrences. To accomplish this task, we can 
apply the following three sub-steps.

\begin{myEnumerate2}
\item First, we transform the system~\eqref{Equ:CoupledDifferenceSys} of order $m$ to a first-order system as follows. For $1\leq i\leq r$, let $n_i\in\NN$ be the maximal value such that $I_i(N+n_i)$ occurs in~\eqref{Equ:CoupledDifferenceSys}. Then one can introduce auxiliary-functions $y_{i,j}(N)$ with $1\leq j< n_{i}$ and $1\leq i\leq r$ where $I_i(N+j)$ is rephrased by $y_{i,j}(N)$ in~\eqref{Equ:CoupledDifferenceSys}. Introducing in addition the relations $y_{i,j}(N+1)=y_{i,j+1}(N)$ and replacing $y_i(N+n_i)$ by $y_{n_i-1}(N+1)$ in~\eqref{Equ:CoupledDifferenceSys} one arrives at a first-order system of difference equations. In other words, solving this first-order system with the unknown $y_{i,j}(N)$ yields immediately the solutions for the $I_i(N)$. For simplicity, we write again $I_i(N)$ instead of $y_i(N)$, and suppose that we are given the system
\begin{equation}\label{Equ:FirstOrderRecN}
\vect{I}(N+1)= A\,\vect{I}(N)+\vect{b}(N),
\end{equation}
where $A$ is an $r\times r$ matrix and $\vect{b}(N)=(b_1(N),\dots,b_r(N))$ is a vector where $b_i(N)$ are expression in terms of nested sums over hypergeometric products.

\item If $A$ is an invertible matrix we skip this step. Otherwise, we proceed as follows. 
 One can perform simple row and column operations in order get again a system 
 of this form with dimension $r'<r$ such that the $r'\times r'$ matrix is invertible and 
such that
we are given linear combinations of the $r'$ unknown integrals which produce the desired integrals 
$I_i(N)$ for $1\leq i\leq r$. Hence, solving this reduced system provides the solution of the original system~\eqref{Equ:FirstOrderRecN}. For further considerations we will continue to work with~\eqref{Equ:FirstOrderRecN} where $A$ is invertible.
 
\item  Finally, we use Z\"urcher's algorithm~\cite{Zuercher:94} that is available in the package~\texttt{OreSys}~\cite{OreSys} in order to uncouple this  system. In general, one obtains $n$ ($1\leq n\leq r$) scalar recurrences, namely, for $1\leq i\leq n$ we get the recurrences
\begin{equation}\label{Equ:SingleREC}
a_{i,i}(N) I_i(N+m_i)+a_{i,m_i-1}(N) I_i(N+m_i-1)+\dots+a_{i,0} I_i(N)=f_i(N)
\end{equation}
of orders $m_i$ where the $a_{i,j}(N)$ are from $\KK(N)$ and the $f_i(N)$ are given as a linear combination of the $b_1(N),\dots,b_r(N)$ over $\KK(N)$ with possible shifts in $N$. Since the $b_i(N)$ are given in terms of nested sums over hypergeometric products, also the $f_i(N)$ can be given in terms of nested sums over hypergeometric products.
In addition, we get explicit linear combinations of the $I_1(N),\dots,I_n(N)$ and the $b_1(N),\dots,b_r(N)$ with possible shifts in $N$ that produce the remaining integrals $I_{n+1}(N),\dots,I_{m}(N)$. More generally, a subset $M$ of $\{I_1,\dots,I_m\}$ can be described by scalar recurrences and the complementary set can be represented by a linear combination of the functions from $M$; after reordering, we may suppose that $M=\{I_1,\dots,I_n\}$.   
\end{myEnumerate2}

\item Finally, we aim at deciding algorithmically if the coefficients $I_i(N)$ of the integrals~\eqref{Equ:HatIAnsatz} can be represented in terms of nested sums over hypergeometric products. In this regard, we have to emphasize that the expressions $b_i(N)$ of $\vect{b}(N)$ arising in~\eqref{Equ:FirstOrderRecN} depend usually on an extra parameter $\ep$ coming from the analytic continuation of the space-time dimension $D=\ep+4$. Similarly, the matrix $A$ in~\eqref{Equ:FirstOrderRecN} depends on this dimensional parameter $\ep$. Thus we usually are given the rational function field $\KK=\KK'(\ep)$ where $\KK'$ is a subfield of $\KK$ containing $\QQ$. Further, there is usually no hope to represent the $b_i(N)$ and also the the integrals $I_i(N)$ in terms of nested sums over hypergeometric products where $\ep$ occurs inside of the sums and products. However, in most applications, there is such a representation, if one considers the functions in its $\ep$-expansion~\cite{BKSS:12,epRepresentation}. More precisely, we assume that the $b_i(N)$ are given in the form
\begin{equation}\label{Equ:biExpansion}
b_i(N)=b_{i,o}(N)\ep^o+b_{i,o+1}(N)\ep^{o+1}+b_{i,o+2}(N)\ep^{o+2}+\dots,
\end{equation}
where the first coefficients $b_{i,j}(N)$ are given in terms of nested sums over hypergeometric 
products (which are free of $\ep$). Further, one is not interested in an expression of the $I_i(N)$, where the nested sums and products depend on $\ep$, but one is interested in the first $l_i$ coefficients of the $\ep$-expansion
$$I_i(N)=I_{i,o}(N)\ep^{o}+I_{i,o+1}(N)\ep^{o+1}+\dots I_{i,o}(N)\ep^{o+l_i},$$
where $I_{i,j}(N)$ is given in terms of nested sums over hypergeometric products (which are free of $\ep$). In our 3-loop calculations we usually have $o=-3$. More precisely, suppose that we are given the initial values 
\begin{equation}\label{Equ:InitialV}
I_i(\nu)=I_{i,o}(\nu)\ep^{o}+I_{i,o+1}(\nu)\ep^{o+1}+\dots I_{i,o+l_i}(\nu)\ep^{o+l_i}+\dots
\end{equation}
for\footnote{Sometimes we need more initial values; the necessary number can be detected during the recurrence solving.} $\nu=1,\dots,m_i$.
Then we can decide algorithmically if the coefficients $I_{i,j}(N)$ can be represented in terms of nested sums over hypergeometric products by executing the following sub-steps.

\begin{myEnumerate2}
\item Note that the $f_i(N)$ for $1\leq i\leq n$ arising in the inhomogeneous parts of the recurrences~\eqref{Equ:SingleREC} are given as a linear combination of the $b_i(N)$. Since they are given in the form~\eqref{Equ:biExpansion} where the $b_{i,j}$ are nested sum expressions over hypergeometric products, one can collect terms in $\ep$ and gets the $\ep$-expansion
\begin{equation}\label{Equ:InhomExpand}
f_i(N)=f_{i,o}(N)\ep^{o}+f_{i,o+1}(N)\ep^{o+1}+\dots f_{i,o}(N)\ep^{o+l_i}+\dots,
\end{equation}
where the $f_{i,j}(N)$ are expressions in terms of nested sums over hypergeometric products (which are free of $\ep$).

\item Given the initial values~\eqref{Equ:InitialV} and the recurrences~\eqref{Equ:SingleREC} with~\eqref{Equ:InhomExpand} where the $f_{i,j}(N)$ are given in terms of nested sums over hypergeometric products, we can activate our recurrence solver~\cite{BKSS:12} based on~\cite{dAlembert,ParticularSol}. 
For all $i$ with $1\leq i\leq n$, we can constructively decide if the integral $I_i(N)$ can be expressed in terms of nested product-sum expressions. If such a representation does not exist for some $i$, we stop and learned that our problem has to be formulated outside of the class of nested sums over hypergeometric products. 

\begin{remark}
The solutions in terms of nested sums over hypergeometric products are also called d'Alembertian solutions~\cite{dAlembert} which can be computed by using algorithms from~\cite{dAlembert,ParticularSol}. Further we remark that the sum representations produced by these algorithms are rather complicated: they are highly nested and the summands have denominators which do not factor nicely. Here we utilize symbolic summation algorithms ~\cite{Summation} based on difference field theory~\cite{DFTheory} and difference ring theory~\cite{DRTheory} in order to rewrite the found solutions in terms of special functions like cyclotomic generalized harmonic sums or nested binomial sums.
\end{remark}

\item We can combine the computed $\ep$-expansions of the $I_1(N),\dots,I_n(N)$ and their shifted versions yielding the $\ep$-expansions of the remaining integrals $I_{n+1}(N),\dots,I_m(N)$ whose coefficients are given in terms of nested sums over hypergeometric products.
\end{myEnumerate2}

\begin{remark}\label{Remark:EliRelation}
Within intermediate steps but also for the final result it is important to get compact representations of the found nested sums and products. In general, we can exploit the underlying summation theory~\cite{DRTheory} in order to compute expressions in terms of nested sums and products where no algebraic (i.e., polynomial) relations exist among the occurring sums and products.
Restricting to harmonic sums and cyclotomic sums, one can employ an alternative and very efficient 
machinery. Namely, it has been shown in~\cite{AIHarmonicSums} that such an optimal representation can be obtained using the underlying quasi-shuffle relations~\cite{Blumlein:2003gb}. Currently, further investigations along the lines of~\cite{AIHarmonicSums} are carried out to avoid the rather involved difference ring algorithms and to reduce the calculations to quasi-shuffle relations and further properties given by the occurring alphabets of the nested sums.
\end{remark}

\item Optionally, we can rewrite the power series solutions in terms of nested integrals involving the extra parameter $x$ using the package \texttt{HarmonicSums}.

\end{myEnumerate}

\noindent The above tactic can be summarized by the following diagram.

$$
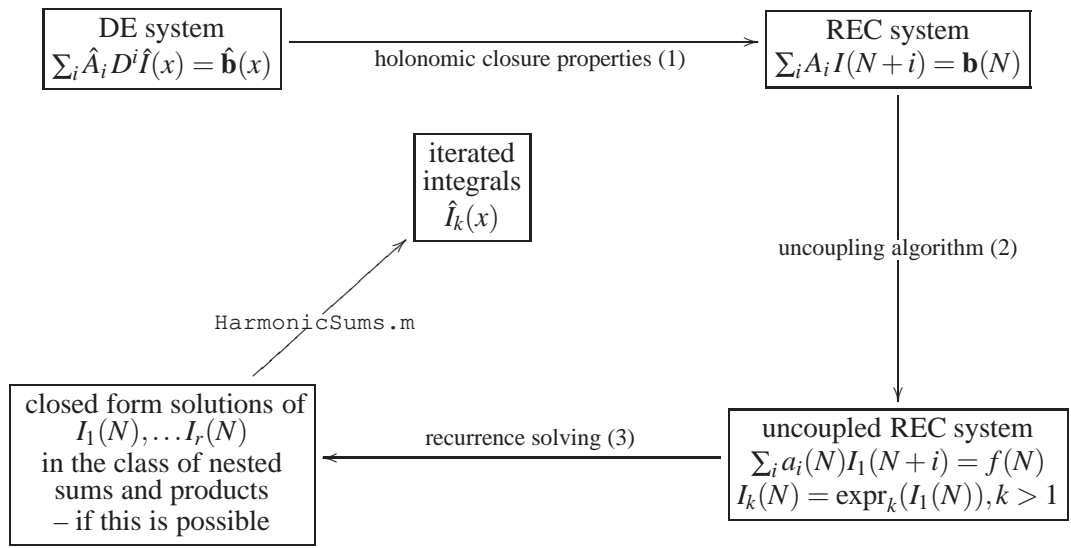
\xymatrix@!R=1.cm@C1.1cm{
\boxed{\txt{DE system\\
$\sum_i\hat{A}_i\,D^i \hat{I}(x)=\hat{\vect{b}}(x)$}{\ar[rrr]_{\txt{\footnotesize holonomic closure properties (1)}}}}&&&{\boxed{\txt{REC system\\
$\sum_iA_i\,I(N+i)=\vect{b}(N)$}}{\ar[ddd]|{\txt{\footnotesize uncoupling algorithm (2)}}}}\\
&{\boxed{\txt{iterated\\integrals\\ $\hat{I}_k(x)$}}}&\\
&&&\\
{\boxed{\txt{
closed form solutions of\\
$I_1(N),\dots I_r(N)$\\
in the class of nested\\
sums and products\\
-- if this is possible}}}{\ar[ruu]|<>(0.4){\txt{\footnotesize\texttt{HarmonicSums.m}}}}
&&&{\boxed{\txt{uncoupled REC system\\
$\sum_i a_i(N)I_1(N+i)=f(N)$\\
$I_k(N)=\text{expr}_k(I_1(N)), k>1$}}{\ar[lll]_{\txt{\footnotesize recurrence solving (3)}}
}}}$$

%-----------------------------------------------------------------------------------------------
\subsection{Tactic 2: Uncoupling a system of differential equations}\label{Sec:DEREC}
%-----------------------------------------------------------------------------------------------

\vspace{1mm}
\noindent
In the above approach we transformed the system of differential equations to a system of recurrences in step~(1), uncoupled the system in step~(2) and finally decided if the unknown integrals can be given in terms of nested sums over hypergeometric products in steps~(3) and~(4). We remark that in step~(1) one obtains usually a coupled system of higher-order difference equations. Uncoupling this system directly, or as proposed in our strategy, bringing it first to a first-order system and uncoupling it afterwards, worked very well for many concrete calculations. However, recently we entered examples where the calculation time of the known uncoupling algorithms blew up dramatically. In contrast to that, the original system provided by IBP methods is usually rather small (and in particular first-order), and the uncoupling algorithms applied to this differential equation system behaves rather tame.

Therefore a different and very promising tactic is to reverse steps~(1) and~(2). More precisely, we can proceed as follows.

\begin{myEnumerate}

\item The uncoupling algorithms available in~\texttt{OreSys}~\cite{OreSys} work for any Ore algebra covering besides the difference case also the differential case. Within \texttt{SolveCoupledSystem} we use again Z\"urcher's algorithm. But this time we uncouple immediately the first-order system~\eqref{Equ:DEProblem} of differential equations. In most instances, the matrix $\hat{A}$ is already invertible. Otherwise, we carry out a similar preprocessing step as worked out in Step~(2b) of Subsection~\ref{Sec:PureREC}.  Summarizing, one obtains $n$ ($1\leq n\leq r$) scalar differential equations of the form
\begin{equation}\label{Equ:DESingle}
a_{i,r_i}(x) D^{r_i}_x\hat{I}_i(x)+a_{i,r_i-1}(x) D^{r_i-1}_x\hat{I}_i(x)+\dots+a_{i,0}(x) D^{0}_x\hat{I}_i(x)=\sum_{j=1}^r\sum_k d_{i,j,k}(x)D_x^k\hat{b}_j(x)
\end{equation}
where the sums on the right hand side are finite and the $a_{i,j}(x)$ and $d_{i,j,k}(x)$ are from $\KK[x]$. In addition, the remaining integrals $\hat{I}_{n+1}(x),\dots,\hat{I}_{m}(x)$ are related to the $\hat{I}_j(x)$  with $1\leq i\leq n$ and the $\hat{b}_i(x)$ with $1\leq i\leq r$ as follows: for all $i$ with $n<i\leq m$ we get
\begin{equation}\label{Equ:RemainingIx}
\hat{I}_i(x)=\sum_{j=1}^n\sum_k\alpha_{i,j,k}(x)D_x^k\hat{I}_{j}(x)+\sum_{j=1}^r\sum_k\beta_{i,j,k}(x)D_x^k\hat{b}_{j}(x)
\end{equation}
where the sums on the right are finite and where the $\alpha_{i,j,k}(x)$ and $\beta_{i,j,k}(x)$ are 
given from $\KK(x)$.

\item Next, we transform the scalar equations~\eqref{Equ:DESingle} to recurrences of the form~\eqref{Equ:SingleREC} with $a_{i,j}(N)\in\KK[N]$ using the same tactic as used in Step~(1) of Subsection~\ref{Sec:PureREC}. Under the assumption that the $b_i(N)$ have an $\ep$-expansion~\eqref{Equ:biExpansion} where the first coefficients $b_{i,j}(N)$ can be given in terms nested sums over hypergeometric products, we can derive $\ep$-expansions~\eqref{Equ:InhomExpand} of the inhomogeneous parts of the recurrences~\eqref{Equ:SingleREC}.

\item Afterwards we apply Step~(3) of Subsection~\ref{Sec:PureREC}. Namely, together with the initial values~\eqref{Equ:InitialV} with $\nu=1,\dots,m_i$
we can decide algorithmically if the coefficients $I_{i,j}(N)$ of~\eqref{Equ:InitialV} for $1\leq i\leq n$ can be represented in terms of nested sums over hypergeometric products.

\item Finally, we compute the coefficients $I_i(N)$ of the integrals~\eqref{Equ:HatIAnsatz} for $n< i\leq m$ as follows.
We replace the $\hat{I}_i(x)$ and $\hat{b}_i(x)$ on the right hand side of~\eqref{Equ:RemainingIx} by their power series representations~\eqref{Equ:HatIAnsatz} and~\eqref{Equ:HatbAnsatz}. In addition, we replace the $I_i(N)$ and $b_i(N)$ arising in the power series representations further by the 
given representation in terms of nested sums over hypergeometric products (possibly in its $\ep$-expansion). 
Then we  utilize the package \texttt{SumProduction}~\cite{Summation} that contains rather efficient algorithms~\cite{VLadders} to calculate the $N$th coefficient of the corresponding power series representation. 
First, we separate the expressions accordingly and get the representation 
\begin{equation}\label{Equ:IExpandX}
\hat{I}_i(x)=\ep^{\omega}\hat{I}_{i,\omega}(x)+\ep^{\omega+1}\hat{I}_{i,\omega+1}(x)+\dots+\ep^{\omega+l_i}\hat{I}_{i,o+l_i}(x)
\end{equation}
for some $\omega\in\ZZ$
where each $\hat{I}_{i,j}(x)$ can be written as a finite sum consisting of summands of the form
\begin{equation}\label{ExtractHCoeff}
\hat{H}(x)=\hat{q}(x)\sum_{N=0}^{\infty}x^N h_1(N)\dots h_l(N).
\end{equation}
Here we have $\hat{q}(x)\in\KK'(x)$ and for all $1\leq i\leq l$ we have that $h_i(N)\in\KK'(x)$, $h_i(N)$ is a hypergeometric product or $h_i(N)$ is a nested sum over hypergeometric products.
With a brute force approach one can now compute the $N$th coefficient for each such expression as given in~\eqref{ExtractHCoeff}, i.e., we can compute an expression $H(N)$ in terms of nested sums over hypergeometric products such that
$$\hat{H}(x)=\sum_{N=0}^{\infty}H(N)x^N$$
holds. This finally yields the $N$th coefficient $I_{i,j}(N)$ of 
$$\hat{I}_{i,j}(x)=\sum_{N=0}^{\infty}I_{i,j}(N)x^N$$ 
in terms of nested sums over hypergeometric products.\\
Namely, first we expand 
$q(x)=\frac{a(x)}{b(x)}$ with $a(x),b(x)\in\KK'[x]$ in a series expansion
%---------------------------------------------------------------------------------------------------------------------------
\begin{equation}\label{Equ:eExp}
\hat{q}(x)=\sum_{N=\mu}^{\infty}q(N)x^i
\end{equation}
%---------------------------------------------------------------------------------------------------------------------------
with $\mu\in\ZZ$ as follows~\cite{VLadders}. Consider the complete factorization 
%---------------------------------------------------------------------------------------------------------------------------
\begin{equation}\label{Equ:RootsOfB}
b(x)=c\,x^{\nu_0}(x-\rho_1)^{\nu_1}(x-\rho_2)^{\nu_2}\dots(x-\rho_v)^{\nu_v}
\end{equation}
%---------------------------------------------------------------------------------------------------------------------------
with $c\in\KK'\{0\}$ and $\rho_i\in\bar{\KK}\setminus\{0\}$ ($\bar{\KK}$ is the algebraic closure of $\KK'$) where $\nu_i\in\NN$ counts the multiplicity of the roots 
$\rho_i$ ($\rho_0=0$). Then, as worked out in~\cite[Thm. 4
1.1]{Stanley:97}, we can calculate the 
expansion
%---------------------------------------------------------------------------------------------------------------------------
$$\frac{1}{b(x)}=\frac{1}{x^{\nu_0}}\sum_{N=0}^{\infty}\beta(N)x^N$$
%---------------------------------------------------------------------------------------------------------------------------
with
%---------------------------------------------------------------------------------------------------------------------------
$$\beta(N)=p_1(N)\,\rho_1^N+p_2(N)\,\rho_2^N+\dots+p_v(N)\,\rho_v^N,$$
%---------------------------------------------------------------------------------------------------------------------------
where the $p_i(N)$ are polynomials in $N$ with degree at most $\nu_i-1$. Now we perform the Cauchy product on $\hat{q}(x)=a(x)\,\frac{1}{x^{\nu_0}}\sum_{N=0}^{\infty}\beta(N)x^N$, and it follows that the coefficient $q(N)$ of the expansion~\eqref{Equ:eExp} can be written again as a linear combination of the $\rho_i^N$ with polynomial coefficients in $N$.
Finally, we obtain
%---------------------------------------------------------------------------------------------------------------------------
$$H(N)=\sum_{k=\mu}^{N+\nu_0}h_1(k)h_2(k)\dots h_l(k)r(k)q(N-k)$$
%---------------------------------------------------------------------------------------------------------------------------
by applying once more the Cauchy product.
Since $q(N-k)$ is given as a linear combination of the $\rho_i^{N-k}=\rho_i^{N}\rho_i^{-k}$ where the coefficients are polynomials in $N$ and $k$,
we can pull out all expressions that depend on $N$. Summarizing, we can write $H(N)$ as an expression in terms of nested sums over hypergeometric products. In particular, the summands of the arising sums are built by the objects 
$h_1(k)h_2(k)\dots h_l(k)$ given in~(\ref{Equ:IExpandX},\ref{ExtractHCoeff}) and the roots 
$\rho_i^k$ from~\eqref{Equ:RootsOfB}.

\end{myEnumerate}

In short, we can summarize the second approach with the following diagram.
$$
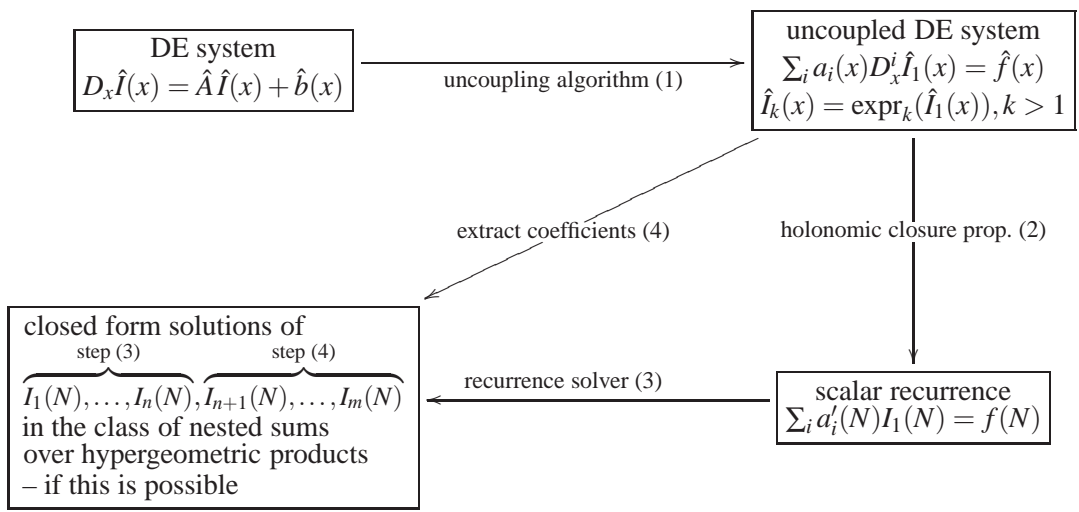
\xymatrix@!R=1.4cm@C2.cm{
\boxed{\txt{DE system\\
$D_x\hat{I}(x)=\hat{A}\,\hat{I}(x)+\hat{b}(x)$}}{\ar[rr]_{\txt{\footnotesize uncoupling algorithm (1)}}}&&
{\boxed{\txt{uncoupled DE system\\
$\sum_i a_i(x)D_x^i \hat{I}_1(x)=\hat{f}(x)$\\
${\bf}\hat{I}_k(x)=\text{expr}_k(\hat{I}_1(x)), k>1$}}{\ar[dd]|{\txt{\footnotesize holonomic closure prop.\ (2)}}}
{\ar[ddll]|{\txt{\footnotesize extract coefficients (4)}}}}\\
\\
{\boxed{\txt{
closed form solutions of\hspace*{1.4cm}\\
$\overbrace{\text{\small$I_1(N),\dots,I_n(N)$}}^{\text{step (3)}},\overbrace{\text{\small$I_{n+1}(N),\dots, I_m(N)$}}^{\text{step (4)}}$\\
in the class of nested sums\hspace*{1cm}\\
over hypergeometric products\hspace*{0.45cm}\\
-- if this is possible\hspace*{2.2cm}}}}
&&{\boxed{\txt{scalar recurrence\\
$\sum_i a'_i(N)I_1(N)=f(N)$}}{\ar[ll]_{\txt{\footnotesize recurrence solver (3)}}}}\\
}$$

%--------------------------------------------------------------------------------------------------
\subsubsection{Improvement 1: compute the $N$th coefficients efficiently (step (4))}\label{Sec:Improvement1}
%--------------------------------------------------------------------------------------------------

\vspace{1mm}
\noindent
Tactic~2 has the advantage that one uncouples the system of differential equations that one is given, e.g., by IBP methods. They are usually in a rather nice shape and can be uncoupled in many examples extremely efficiently. However, this big advantage is paid by the challenge to compute the $N$th coefficient of the $\hat{I}_{i,j}(x)$ in~\eqref{Equ:IExpandX}: IBP methods usually calculate expressions with rather complicated denominators, i.e., one obtains subexpressions $\hat{H}(x)$ in~\eqref{ExtractHCoeff} where the rational functions $\hat{q}(x)=\frac{a(x)}{b(x)}$ have denominators $b(x)$ that do not have nice irreducible factors over $\KK'$ (mostly over $\QQ$). As a consequence, the method proposed in step~(4) yields alien sums whose summands are not expected to appear within the final result. 

\medskip

\noindent\textit{Example.} Consider 
$$\hat{H}(x)=\frac{1}{1-a x}\sum_{N=0}^{\infty } x^N S_{2,1}({N})=\sum_{N=0}^{\infty} H(N)x^N$$
with $a\in\ZZ$. Then the $N$th coefficient is
$$
H(N)=\begin{cases}
\frac{a^{N+1}}{a-1}S_{2,1}\big({{\frac{1}{a},1},N}\big)
-\frac{1}{a-1}S_{2,1}({N})&\text{ if }a\in\ZZ\setminus\{0,1\}\\
S_{2,1}({N})&\text{ if }a=0\\
(N+1)S_{2,1}(N)-\frac12 S_2(N)-\frac12 S_1(N)^2&\text{ if }a=1.
\end{cases}
$$
The arising sums with $a=0,\pm1$ pop up almost everywhere within our 
calculations~\cite{DiffCalculations}. So denominators of the form $1-ax$ with $a=0,\pm1$ are no surprise. In some instances, also the generalized sums with $a=\pm 2$ arise, and thus also the underlying denominators will appear. However, sums coming from $a\in\ZZ$ and $a\neq0,\pm1,\pm 2$ 
have not occurred in our ongoing calculations, but will arise within intermediate calculations when one executes step~(4). The situation gets worse if non-linear factors are treated. One of the simplest cases is $\hat{q}(x)=\frac1{1-x-x^2}$. In this situation, e.g., the subexpression
$$\frac{1}{1-x-x^2}\sum_{N=0}^{\infty} x^N S_{2,1}({N})=\sum_{N=0}^{\infty} H(N)x^N$$
has the $N$th coefficient
\begin{equation}\label{Equ:AlienSums}
H(N)=\frac{3 \sqrt{5}-5}{10}(-1)^N \Phi^{-N-1} S_{2,1}({{-\Phi ,1},N})
        +\frac{5+3 \sqrt{5}}{10} \phi ^{-N-1} S_{2,1}({{\phi,1},N})
        -S_{2,1}({N})
\end{equation}
with the golden ration $\Phi=\frac{\sqrt{5}+1}{2}$ and $\phi=\frac{\sqrt{5}-1}{2}$.

\medskip

One option is to produce the final expression with all these artificial sums. Then eliminating all relations among these sums (see Remark~\ref{Remark:EliRelation}) will lead to an expression where all alien sums will collapse and the expected sums will remain. However, in many calculations the number of sums in the summands~\eqref{ExtractHCoeff} can be quite large -- we considered cases with up to 1000 sums. Performing than step~(4) naively the number of sums will explode. Even worse, if algebraic numbers like $\phi$ and $\Phi$ arise, the computation of algebraic relations 
turns into a real computer algebra challenge.\\
In order to avoid these troubles, the following alternative option worked out in~\cite{VLadders} has been incorporated into the package~\texttt{SumProduction} that we use heavily as a subroutine within our newly developed package~\texttt{SolveCoupledSystem}. Recall that the desired coefficients $\hat{I}_{i,j}(x)$ of~\eqref{Equ:IExpandX}
can be written as a big expressions summed up by subexpressions of the form~\eqref{ExtractHCoeff}. We truncate now the infinite sums within the given expression of $\hat{I}_{i,j}(x)$, i.e., the arising sums are of the form 
$$\sum_{N=0}^{A}x^N h_1(N)\dots h_l(N)$$
(instead of the form~\eqref{ExtractHCoeff}). Now we eliminate all algebraic relations among these 
nested sums over hypergeometric products (involving in addition the parameter $x$). Here the 
following magic happens: in all our examples, we observed that all sums that would contribute to 
alien terms as given in~\eqref{Equ:AlienSums} vanish. 

Still, the proposed tactic is rather expensive to treat all truncated sums (all with the same upper bound $A$) simultaneously. As 
a compromise we filter out only those subexpressions that might contribute to alien sums, more 
precisely, we partition the subexpressions into several parts:
(a) sums with denominators that have only nice irreducible factors (over $\QQ$) in the 
denominator, 
which we keep untouched (this is usually the largest part of the full expression), and (b) sums 
with denominators that have only bad irreducible factors in the denominator, i.e., which are of 
the 
form $(1-ax)$ with $a\neq0,\pm1$ (or even $a\neq\pm2$) and factors which are not linear (over $\QQ$). 
Even more, we partition the sums with bad denominators further such that sums are collected which 
have common bad factors. Then we compute for each such partition of bad sums, which consists of conquerable subexpressions, all algebraic relations. In all our examples all these bad sums within their clusters vanish. Finally, we compute the limit $A\to\infty$ and end up at an alternative expression of $\hat{I}_{i,j}(x)$ where now all the unwanted sums are gone. Computing finally the $N$th coefficient as worked out in step~(4) will lead to an expression where the nested sums over hypergeometric products have nice denominators. In particular, the number of these sums is now manageable, and we can compute an alternative representation in terms of nested sums where all arising sums are algebraically independent among each other (see Remark~\ref{Remark:EliRelation}).

%-----------------------------------------------------------------------------------------------------
\subsubsection{Improvement 2: compute recurrence relations of smaller order (step 
(2))}\label{Sec:Improvement2}
%-----------------------------------------------------------------------------------------------------

\vspace{1mm}
\noindent
Recall the first steps of our proposed procedure.
Suppose we obtained the scalar differential equation~\eqref{Equ:DESingle} in step~(1) with $a_{i,j}(x)$ and $d_{i,j,k}(x)$ being from $\KK[x]$. Then plugging in~\eqref{Equ:HatIAnsatz} into~\eqref{Equ:DESingle}, using the rule~\eqref{Equ:DRelation} and taking the $N$th coefficient, we end up at a linear recurrence of the form~\eqref{Equ:CoupledDifferenceSys} where the
the $f_i(N)$ are given as a linear combination of the $b_1(N),\dots,b_r(N)$ over $\KK(N)$ with possible shifts in $N$.
Under the assumption that the $b_i(N)$ have an $\ep$-expansion~\eqref{Equ:biExpansion} where the first coefficients $b_{i,j}(N)$ can be given in terms nested sums over hypergeometric products, we can derive an $\ep$-expansion~\eqref{Equ:InhomExpand} of the inhomogeneous part of the recurrence~\eqref{Equ:SingleREC}. 

Observe that the order of the derived recurrence is bounded by the maximum of the degrees of the coefficients $a_{i,j}(x)$. 
If the order is not too large, the method proposed above works perfectly fine. However, in recent examples we calculated recurrences of order 16 or higher, and it is then almost impossible to calculate 16 (or more) initial values that are needed for step~(3). Luckily, in all these examples it turns out that $d(x)=\gcd(a_{i,0}(x),\dots,\gcd(a_{i,r_i}(x))\in\KK[x]$ has a rather high degree. Hence dividing~\eqref{Equ:DESingle} through $d(x)$ leads to the differential equation
\begin{equation}\label{Equ:DESingleMod}
a'_{i,r_i}(x) D^{r_i}_x\hat{I}_i(x)+a'_{i,r_i-1}(x) D^{r_i-1}_x\hat{I}_i(x)+\dots+a'_{i,0}(x) D^{0}_x\hat{I}_i(x)=\sum_{j=1}^r\sum_k d'_{i,j,k}(x)D_x^k\hat{b}_j(x)
\end{equation}
where the $a'_{i,j}=\frac{a_{i,j}(x)}{d(x)}\in\KK[x]$ have substantially smaller degrees and where $d'_{i,j,k}=\frac{d_{i,j,k}}{d(x)}\in\KK(x)$. Hence taking the $N$th coefficient on the left hand side of~\eqref{Equ:DESingleMod} will yield a difference operator in $I_i(N)$ whose order is substantially smaller (the order is bounded by the maximum of the degrees of the coefficients $a'_{i,j}(x)$). However, in order to get the $N$th coefficient on the right hand side of~\eqref{Equ:DESingleMod}, i.e., in order to get a recurrence of the from~\eqref{Equ:SingleREC}, further calculations are necessary.
Since $d'_{i,j,k}\in\KK(x)$ is usually not a polynomial in $x$, we have to apply again the rather 
involved calculations steps as sketched in step~(4). In this regard, the improvements of Section~\ref{Sec:Improvement1} play a central role to carry out these calculations efficiently and to provide a linear recurrence with a substantially smaller recurrence  order for step~(2) of our procedure.

%------------------------------------------------------------------------------------------------------
\section{Conclusion}\label{Sec:Conclusion}
%------------------------------------------------------------------------------------------------------

\vspace{1mm}
\noindent
In both algorithms, presented in Subsections~\ref{Sec:PureREC} and~\ref{Sec:DEREC} respectively, 
one ends up at scalar recurrences~\eqref{Equ:SingleREC}: sometimes they are the same, sometimes 
one or the other method finds a better recurrence (with smaller coefficient size or with lower 
recurrence order). This suggests to apply both tactics (up to a certain point) and to execute the 
version in full detail that is more appropriate for the concrete problem. As mentioned already 
above, a central advantage of the second tactic is that one can uncouple the system straightforwardly 
(without any preprocessing steps as sketched in Subsection~\ref{Sec:PureREC} that might blow up the 
system). In some instances this leads to a much better space--time behavior. However, in this approach 
one has to compute the $N$th coefficient of the remaining integrals, which again can be rather time 
consuming. But using our sophisticated symbolic summation technologies 
(see Subsection~\ref{Sec:Improvement1}) this problem turns out to be feasible in many examples. The 
second advantage of our new method is that one might find recurrences with smaller orders 
(see Subsection~\ref{Sec:Improvement2}). As a consequence, one needs less initial values to 
determine the respective master integrals. Since the calculations of such initial values is 
rather challenging, we expect that this last feature will support future calculations.

%--------------------------------------------------------------------------------------------------

%-----------------------------------------------------------------------------

\end{document}